\documentclass[conference]{IEEEtran}
\IEEEoverridecommandlockouts
\usepackage{cite}
\usepackage{amsmath,amssymb,amsfonts}
\usepackage{algorithmic}
\usepackage{graphicx}
\usepackage{textcomp}
\usepackage{xcolor}
\def\BibTeX{{\rm B\kern-.05em{\sc i\kern-.025em b}\kern-.08em
    T\kern-.1667em\lower.7ex\hbox{E}\kern-.125emX}}
\usepackage{multirow}
\usepackage{listings}
\usepackage{booktabs}
\usepackage{graphicx}
\usepackage{makecell}
\usepackage{hyperref}
\usepackage{tcolorbox}
\usepackage{soul}
\usepackage{colortbl, xcolor}

\definecolor{mistyrose}{rgb}{1.0, 0.89, 0.88}
\definecolor{pastelpink}{rgb}{1.0, 0.82, 0.86}
\definecolor{palepink}{rgb}{0.98, 0.85, 0.87}
\definecolor{seashell}{rgb}{1.0, 0.96, 0.93}
\definecolor{snow}{rgb}{1.0, 0.98, 0.98}

\newcommand{\rqone}{RQ1: How accurately can regular expressions detect code constructs in textbooks?}
\newcommand{\rqtwo}{RQ2: What Python code constructs have been found in the textbooks?}
\newcommand{\rqthree}{RQ3: To what extent do the sequences in which Python code constructs are introduced in textbooks agree with the competency levels?}
\newcommand{\rqfour}{RQ4: For which Python constructs is there major disagreement between where they have been introduced in textbooks and their competency level?}
\newcommand{\rqfive}{RQ5: Why do those disagreements (identified in RQ4) happen?}

\definecolor{applegreen}{rgb}{0.55, 0.71, 0.0}
\definecolor{chestnut}{rgb}{0.8, 0.36, 0.36}
\definecolor{gray(x11gray)}{rgb}{0.75, 0.75, 0.75}
\definecolor{mygray}{RGB}{128,128,128}

\begin{document}

\title{Towards Identifying Code Proficiency through \\the Analysis of Python Textbooks}

\author{
\IEEEauthorblockN{Ruksit Rojpaisarnkit, Gregorio Robles*, Raula Gaikovina Kula \\ Dong Wang$^{\dagger}$, Chaiyong Ragkhitwetsagul$^{\star}$, Jesus M. Gonzalez-Barahona*, Kenichi Matsumoto}
\textit{Nara Institute of Science and Technology, Japan}\\
\textit{Universidad Rey Juan Carlos, Spain*}\\
\textit{Tianjin University, China$^{\dagger}$}\\
\textit{Faculty of ICT, Mahidol University, Thailand$^{\star}$}\\

\{rojpaisarnkit.ruksit.rn1, raula-k, matumoto\}@is.naist.jp \\ grex@gsyc.urjc.es, dong\_w@tju.edu.cn, chaiyong.rag@mahidol.edu, jgb@gsyc.es

}

\maketitle

\begin{abstract}
Python, one of the most prevalent programming languages today, is widely utilized in various domains, including web development, data science, machine learning, and DevOps. 
Recent scholarly efforts have proposed a methodology to assess Python competence levels, similar to how proficiency in natural languages is evaluated. 
This method involves assigning levels of competence to Python constructs—for instance, placing simple 'print' statements at the most basic level and abstract base classes at the most advanced.
The aim is to gauge the level of proficiency a developer must have to understand a piece of source code. 
This is particularly crucial for software maintenance and evolution tasks, such as debugging or adding new features. 
For example, in a code review process, this method could determine the competence level required for reviewers.
However, categorizing Python constructs by proficiency levels poses significant challenges. 
Prior attempts, which relied heavily on expert opinions and developer surveys, have led to considerable discrepancies.
In response, this paper presents a new approach to identifying Python competency levels through the systematic analysis of introductory Python programming textbooks. 
By comparing the sequence in which Python constructs are introduced in these textbooks with the current state of the art, we have uncovered notable discrepancies in the order of introduction of Python constructs.
Our study underscores a misalignment in the sequences, demonstrating that pinpointing proficiency levels is not trivial.
Insights from the study serve as pivotal steps toward reinforcing the idea that textbooks serve as a valuable source for evaluating developers' proficiency, and particularly in terms of their ability to undertake maintenance and evolution tasks.
\end{abstract}


\section{Introduction}
Proficiency in programming languages, especially Python, is crucial for software developers, not just for creating software but also for its maintenance and evolution. The concept of proficiency levels\footnote{The terms "proficiency" and "competency" are used interchangeably in this paper. Competency refers to the essential skills required, while proficiency implies a certain mastery of them.} in programming is pivotal, as it directly impacts the efficiency of software development tasks, including code reviews. In such processes, being able to determine the competence level necessary to understand a piece of source code is essential. This determination allows for the assignment of code review tasks to developers with appropriate levels of competence, thereby facilitating smoother onboarding onto software projects as noted by Steinmacher et al.~\cite{steinmacher2015systematic}.
Moreover, in situations involving bug fixes, where the section of the code containing the error has been identified, the selection of the developer for the task usually favors those with the most file-specific experience~\cite{anvik2006should, etemadi2022task}. However, the introduction of an additional evaluative factor—the necessary competence level to implement the change—could revolutionize this assignment process. It opens the possibility of delegating tasks that require a lower level of expertise to less experienced developers, thereby enabling more seasoned developers to focus on challenges that match their higher skill set. This approach not only maximizes the efficient use of resources but also supports the growth and development of novice programmers within the team~\cite{chouchen2021whoreview}.

Even in an era where AI technologies such as Large Language Models (LLMs) have the potential to significantly streamline the coding process, the fundamental understanding of source code remains indispensable for software engineers~\cite{liang2024}. While LLMs can enhance process efficiency, a deep understanding of the code is crucial for developers to effectively navigate the software development cycle and sustain creativity. Furthermore, as the concept of \emph{responsibility} is unequivocally tied to humans—and by extension, to developers—the role of software engineers in ensuring the proper functioning of software systems becomes even more critical~\cite{schieferdecker2020responsible}. The ability to understand and interpret the source code is essential in fulfilling this responsibility~\cite{gopstein2020thinking,lu2022towards}.
Modern software projects often incorporate multiple programming languages, making mastery over all of them a challenging endeavor~\cite{vasilescu2013babel,li2024multilingual}. Ensuring that the body of code aligns with the competency level of a software team is therefore crucial. The emergence of code, whether generated by an LLM or a developer, that is not comprehensible to any team member, poses a significant risk. Such scenarios underscore the ongoing necessity for developers to maintain a robust understanding of source code, reinforcing the value of coding proficiency in the face of advancing LLM technologies.

The absence of a universally acknowledged framework to gauge a developer's proficiency in programming languages starkly contrasts with the structured assessment methodologies available for natural languages, such as the widely adopted Common European Framework of Reference for Languages (CEFR)~\cite{north2014cefr}. This framework categorizes language proficiency into three primary levels—A (basic), B (independent), and C (proficient)—each further divided to reflect different degrees of skill~\cite{jones2009european}.
In the realm of programming, efforts to bridge this gap have led to the adaptation of the CEFR framework for languages like Python. A significant development in this direction is \textit{pycefr}, a tool that evaluates Python code to ascertain the proficiency level a programmer needs for comprehension~\cite{Robles2022}. 
By analyzing Python constructs for complexity and mapping them to CEFR-equivalent levels, \textit{pycefr} aims to standardize programming proficiency assessment. 
This initiative, drawing inspiration from the pedagogical sequence in ``Learning Python" by Mark Lutz ~\cite{lutz2001programming} and validated through expert feedback, offers a detailed mapping of Python constructs to proficiency levels, from basic 'print' statements to advanced concepts like list comprehensions and the 'zip' function. 
Nonetheless, even the authors of \textit{pycefr} acknowledge the ongoing potential for discussion regarding the precise assignment of these constructs to the different levels, indicating an openness to evolving interpretations in the pursuit of standardizing programming language proficiency assessment.

In this study, we aim to conduct a systematic evaluation of the competency assignment of Python constructs as delineated by \textit{pycefr}. We hypothesize that there exists a universal pedagogical sequence in the introduction of programming concepts, beginning with the simplest and gradually escalating to more complex ones. This premise suggests that introductory programming textbooks are designed with a deliberate structure, methodically advancing from foundational to advanced concepts in a consistent and incremental fashion. 
In the past, version control systems, mailing lists, bug-tracking systems, package repositories, Q\&A sites, and blogs have been mined~\cite{amann2015software}. However, all these sources lack a deliberate ordering that reflects the authors' intentions regarding the sequence in which a programming language should be learned.

To verify the presence of such a sequence in Python learning materials, we undertook a systematic and extensive review of introductory Python textbooks. Our approach involved selecting a range of textbooks geared towards Python beginners and analyzing the point at which various \textit{pycefr}-identified Python constructs are introduced.
We focus particularly on constructs that deviate significantly from the recommended sequence by \textit{pycefr}, highlighting instances where constructs appear in unexpected positions within the textbooks. These discrepancies, once identified, will be subjected to a detailed qualitative and individualized examination to uncover the underlying reasons for the divergence.

 Key findings indicate that the assignment of proficiency levels is not trivial compared to natural language. 
 Despite a significant alignment (around 80\%) between the pycefr assignments and textbook sequences, disparities are more pronounced for complex constructs categorized under levels B2, C1, and C2. 
 This divergence indicates a more universally agreed sequencing for basic constructs compared to their advanced counterparts.
  The recommendation to reposition constructs traditionally introduced at levels A2 and B1, such as the simple while loop and function imports, to the more basic A1 level, underscores their fundamental role and widespread presence in the initial stages of programming textbooks. 
The study points to a need for a reassessment of pycefr classifications, advocating for adjustments in the assignments performed.
However, we also have identifed in Python textbooks constructs that can be considered \emph{out of order}, pointing out to the need to recognize and filter those out if the method proposed in this paper is to be used to obtain a \emph{complete} assigment of Python constructs to competency levels. 

\section{Related Work}
\label{sec:related}

\textbf{Experience and Skillsets.}
There are several studies focusing on understanding and measuring the developer's expertise or proficiency in OSS communities, such as the proposal of the H-index to help the ``career advancement'' of
developers~\cite{capiluppi2012developing}.
\cite{vadlamani2020studying} revealed that an expert would necessarily have to possess adequate soft skills such as effective and clear communication, and analytical thinking.
It has been shown that prior language proficiency and experience were factors that influence developer onboarding~\cite{casalnuovo2015developer}.
\cite{dey2021representation} studied the representation of developer expertise, aiming to increase the trust and efficiency of OSS ecosystems at large.
Several models have been proposed to recommend expertise.
To name a few, \cite{venkataramani2013discovery} built a recommendation system for StackOverflow by ranking the expertise of the developers in a target domain from GitHub.
\cite{liang2022understanding} developed a model of skills in OSS that considers the many contexts and they found that OSS contributors are actively motivated to improve skills.
\cite{atzberger2022codecv} proposed an approach namely \textit{CodeCV} to mine the expertise of GitHub users
from coding activities (i.e., commits).

\textbf{Pythonic Coding}.
A group of studies has been carried out to investigate Pythonicity, deemed acceptable and proficient (competent) code from the Python community.
Pythonic code follows guiding principles and practices within the Python community.
\cite{alexandru2018usage} performed groundwork toward understanding Python idioms in actual source code and they observed that writing
Pythonic code can be an indicating factor of expertise of Python developers.
\cite{sakulniwat2019visualizing} visualized the usage of the \textit{with open} idiom in software projects and results showed that developers tend to adopt this idiom over time.
Phan-udom et al.~\cite{phan2020teddy} introduced an automated tool called Teddy that suggests the use of Pythonic idioms based on code changes in GitHub pull requests.
\cite{leelaprute2022does} analyzed the performance of Pythonic codes at scale.
They revealed that writing in Pythonic idioms
may save memory and time.
To support the automatic task of refactoring non-idiomatic Python code, \cite{zhang2022making} developed an automatic tool that detects nine types of non-idiomatic codes, yielding high accuracy.
Zhang et al.~\cite{Zhang2023} performed a large-scale study of nine Pythonic idioms in terms of performance, i.e., execution time. 
They observed that the Pythonic idioms can either speed up or slow down the performance.

\textbf{Computing Python Code Competency.} 
\textit{pycefr}~\cite{Robles2022} is an automated tool that analyzes Python code and measures the proficiency level required to understand a fragment of Python code. 
The tool adopts the concept of the Common European Framework of Reference for Languages (CEFR), which is widely used in natural languages, to categorize Python code fragments into six proficiency levels (A1, A2, B1, B2, C1, C2)~\cite{wisniewski2017empirical}. 
Levels A1 and A2 include code constructs that are considered to be for basic users of the Python language such as \textit{print}, \textit{if} statements, or nested lists. Levels B1 and B2 include code constructs that are for independent Python users such as lists with a dictionary, \textit{with} statement, or list comprehensions. Levels C1 and C2 are for code constructs that are used by proficient users of the Python language such as generator functions, or meta-classes.
\textit{pycefr} accepts a directory of a Python project as an input and creates a report of pairs of code constructs and their respective code competency levels. The assignments of code constructs to proficiency levels in \textit{pycefr} can be easily changed in the configuration file of the tool, without having to modify the tool's source code.
Initial work has begun on evaluating the usefulness of \textit{pycefr}. For instance, Febriyanti et al.~\cite{Febriyanti2022} explore the \textit{pycefr} tool to study the contributor's code competency in four PyPI library projects. They found that code in most files contains basic-level proficiency. By using visualizations, they also found that most contributors contribute code that is mostly categorized into basic levels.

\section{Goal, Method, and RQs}
\label{sec:method-goal-rqs}

\begin{figure*}
    \centering
    \includegraphics[width=.9\textwidth]{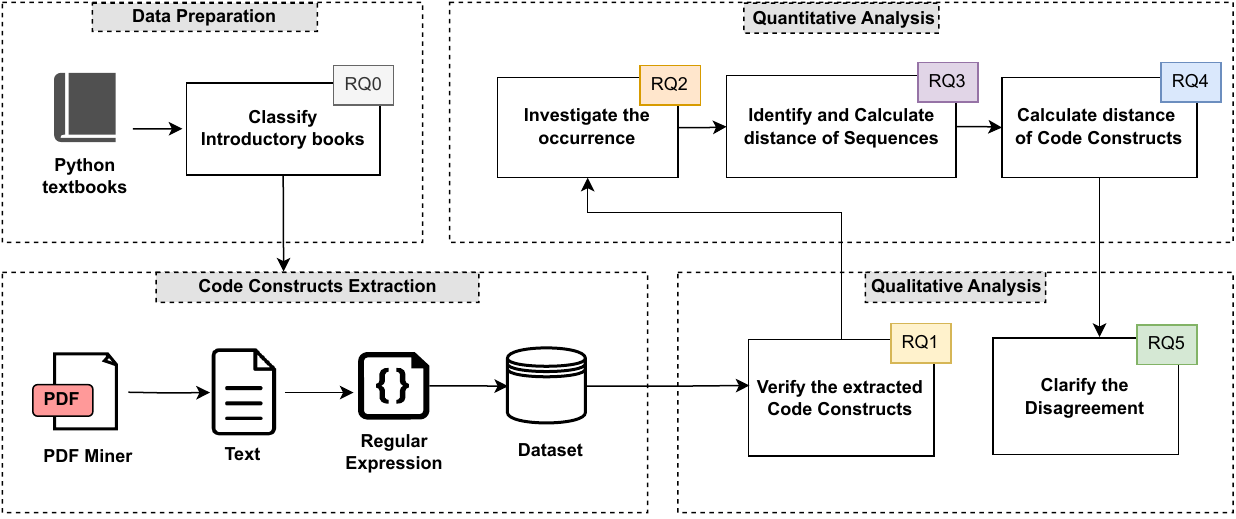}
    
    \caption{Research Architecture}
    \label{fig:methodology}
\end{figure*}

The main goal of the paper is to evaluate if the sequence in which Python constructs are presented in introductory Python textbooks agrees with the competency levels that have been proposed with the \textit{pycefr} tool. 

Figure~\ref{fig:methodology} offers an overview of the research process we have followed. The process draws upon an empirical software engineering approach that combines quantitative and qualitative elements, mining the rich data source of programming textbooks.
Therefore, first, we retrieved electronic versions of Python books available in our university libraries (we searched in their databases for the term "Python" in the title or in the tags), resulting in a list of 80 electronic books in PDF format. 
Then, we removed false positives (e.g., books that had no relation with the programming language), duplicates, and books for which we had several editions, and obtained a list of 28 books that could be candidates for introductory books to programming using Python.
Next, four of the seven authors manually inspected the 28 books and determined those that we considered to be introductory to programming, The evaluation resulted in an overall agreement of 85.12\%, with free-marginal kappa values of 0.70 and a 95\% CI for free-marginal kappa between [0.52, 0.88], which are quite high agreement values. In particular, for 20 of the 28 books, there was unanimity in the decision; 10 of them were classified as ``Beginner'' and 10 as ``Non-beginner'' by the five raters. Of the other 8 books, for 6 there was only one dissenting evaluator; in two cases all but one had evaluated it as ``Beginner'', and in the other four cases all but one had indicated ``Non-beginner''. In the other 2 books, the discrepancy was 3 to 2. In any case, all situations were discussed, reaching an agreement. The final list is composed of 12 textbooks (see Table~\ref{tab:book-list}).

\begin{table}
    \centering
    \caption{Examples of regular expression used to identify Python constructs}
    \label{tab:regex-table}
    \begin{tabular}{ll}
    \toprule
    Python Construct & Regular Expression \\
    \midrule
    printfunc & \texttt{print\textbackslash{}(.*\textbackslash{}n.*\textbackslash{})} \\
    simplelist & \texttt{\textbackslash{}w+\textbackslash{}s*=\textbackslash{}s*[\textbackslash{}s*.*\textbackslash{}s*]} \\
    fornested & \texttt{for\textbackslash{}s+\textbackslash{}w+.*\textbackslash{}s+in\textbackslash{}s+\textbackslash{}w+.*:[\textbackslash{}s\textbackslash{}S]+}
    \\
    & \texttt{for\textbackslash{}s+\textbackslash{}w+.*\textbackslash{}s+in\textbackslash{}s+\textbackslash{}w+.*} \\
    whilecontinue & \texttt{while\textbackslash{}s+.*:[\textbackslash{}s\textbackslash{}S]+if\textbackslash{}s+.*:[\textbackslash{}s\textbackslash{}S]} \\
    & \texttt{+continue}\\
    zipfunc & \texttt{zip\textbackslash{}(.*\textbackslash{})}\\
    \bottomrule
    \end{tabular}
\end{table}

Our extraction process consists of three main steps. Firstly, we utilize the \textit{PDF Miner} tool to convert textbooks into text format, allowing us to extract and parse the necessary data. Secondly, we develop a set of regular expressions designed to identify 94 Python constructs, as defined by \textit{pycefr}. Table \ref{tab:regex-table} provides examples of the regular expressions and corresponding Python constructs used in this process. Lastly, we store the results in a JSON file, which includes all the identified Python constructs from each page of the textbook, facilitating easy querying and analysis of the data. In order to achieve our goal, we have derived the following research questions:

\noindent\textbf{\rqone}\\
\noindent\textit{\underline{Goal:}} Given that we extract Python code constructs using regular expressions, we aim to ascertain how good our approach is. We will therefore manually analyze a representative sample of code snippets.\\
\noindent\textit{\underline{Outcome:}} We expect to obtain a high precision and recall, to offer high confidence in our results.

\noindent\textbf{\rqtwo}

\noindent\textit{\underline{Goal:}} Our aim is to investigate the appearance of each code construct in the introductory books to Python.

\noindent\textit{\underline{Outcome:}} We will quantify how many of the 94 code constructs defined by \textit{pycefr} appear in the textbooks. Thus, we will provide the distribution of the number of appearances of code constructs in the books. We will analyze the distribution, based on the competency level, to see if basic constructs (e.g., A1 and A2) appear more frequently than complex ones. We will also provide and discuss the list of those constructs that appear in all books and the list of those that do not appear in any of them.

\noindent\textbf{\rqthree}

\noindent\textit{\underline{Goal:}} We aim to find out how to align textbooks to a \emph{common} sequence.  Therefore, we analyze when code constructs appear in the Python introductory books for the first time. Then, we will use \textit{pycefr}'s assignment of constructs to competency levels, and analyze how much the sequences in the books have in common with it.

\noindent\textit{\underline{Outcome:}} First, we will analyze the alignment visually, by grouping code constructs by their competency level and first appearance. Then, we will use a distance metric to quantitatively the alignment of textbooks and \textit{pycefr}.

\noindent\textbf{\rqfour}\\
\noindent\textit{\underline{Goal:}}
We aim to identify those Python code constructs where there is a major divergence between their location in textbooks and their \textit{pycefr} competency level. 

\noindent\textit{\underline{Outcome:}} For every Python code construct, we will provide a distance measure of where we found it in textbooks and what level it has been assigned by \textit{pycefr}. As not all constructs appear in all books, we will provide this measure in absolute and relative terms.

\noindent\textbf{\rqfive}

\noindent\textit{\underline{Goal:}} We want to manually evaluate those Python code constructs from RQ4 that have shown to have a high divergence between the competency level proposed by \textit{pycefr} and its location in the Python textbooks.

\noindent\textit{\underline{Outcome:}} We will investigate the code snippets where code constructs appear for the first time in Python textbooks, analyze their location, and their context, and based on this evaluate them according to their coherence.

\section{Findings}
\label{sec:findings}

% \subsection{\rqzero}

\begin{table}[]
\caption{List of Python textbooks analyzed}
\label{tab:book-list}
\begin{tabular}{clc}
 \toprule
\textbf{No.} & \textbf{Book Name} & \textbf{Ref.}                                                                 \\ \toprule
1   & A Beginner's Guide To Python 3 Programming &\cite{hunt2019beginners}                        \\
2   & Programming in Python 3 & \cite{summerfield2010programming} \\
3   & Learning Python, 5\textsuperscript{th} edition       & \cite{lutz2013learning}                                      \\
4   & Python 3 for Absolute Beginners   & \cite{hall2010python}                                       \\
5   & Head First Python             & \cite{barry2016head}                                           \\
6   & Python Crash Course        & \cite{matthes2015python}                                              \\
7   & Core Python Programming, 2\textsuperscript{nd} edition  & \cite{chun2006core}                         \\
8   & Python for Kids                  & \cite{briggs2012python}                                        \\
9   & Python Projects for Kids         & \cite{ingrassellino2016python}                                       \\
10  & Making Use of Python   & \cite{gupta2003making}                                                  \\
11  & A Python Book: Beginning Python, ... & \cite{kuhlman2011python}   \\
12  & Think Python, 2\textsuperscript{nd} Edition  & \cite{downey2015think}         \\\bottomrule                                    
\end{tabular}

\vspace{-8pt}

\end{table}

\subsection{\rqone}

With this research question, we aim to evaluate the regular expressions used to mine Python constructs from the textbooks by conducting a manual verification. We extracted a total of 31,333 code constructs from the 12 introductory textbooks. To achieve a 95\% confidence level, we had to manually analyze a sample of 380 code constructs.  The validation is based on two aspects: i) whether the extracted text is code or not, and ii) whether the code is classified as the correct code construct or not. 

Our results reveal that 297 out of 380 constructs have been i) correctly extracted as a code and ii) matched their code construct type. 19 samples have been classified as code, but as the wrong code construct; and 64  samples were not code (e.g., a \textit{print} in the text that does not relate to the \textit{print} Python method). Consequently, the accuracy, precision, and recall of our regular expression extraction procedure are 83.16\%, 82.27\%, and 93.99\%, respectively.

\subsection{\rqtwo}

%%%%%%%%%%%%%%%%%%%%%%%%%%%%%%%%%%%%%%%%%%%%%%%%
\begin{figure*}
  \centering
  \includegraphics[width=.9\textwidth]{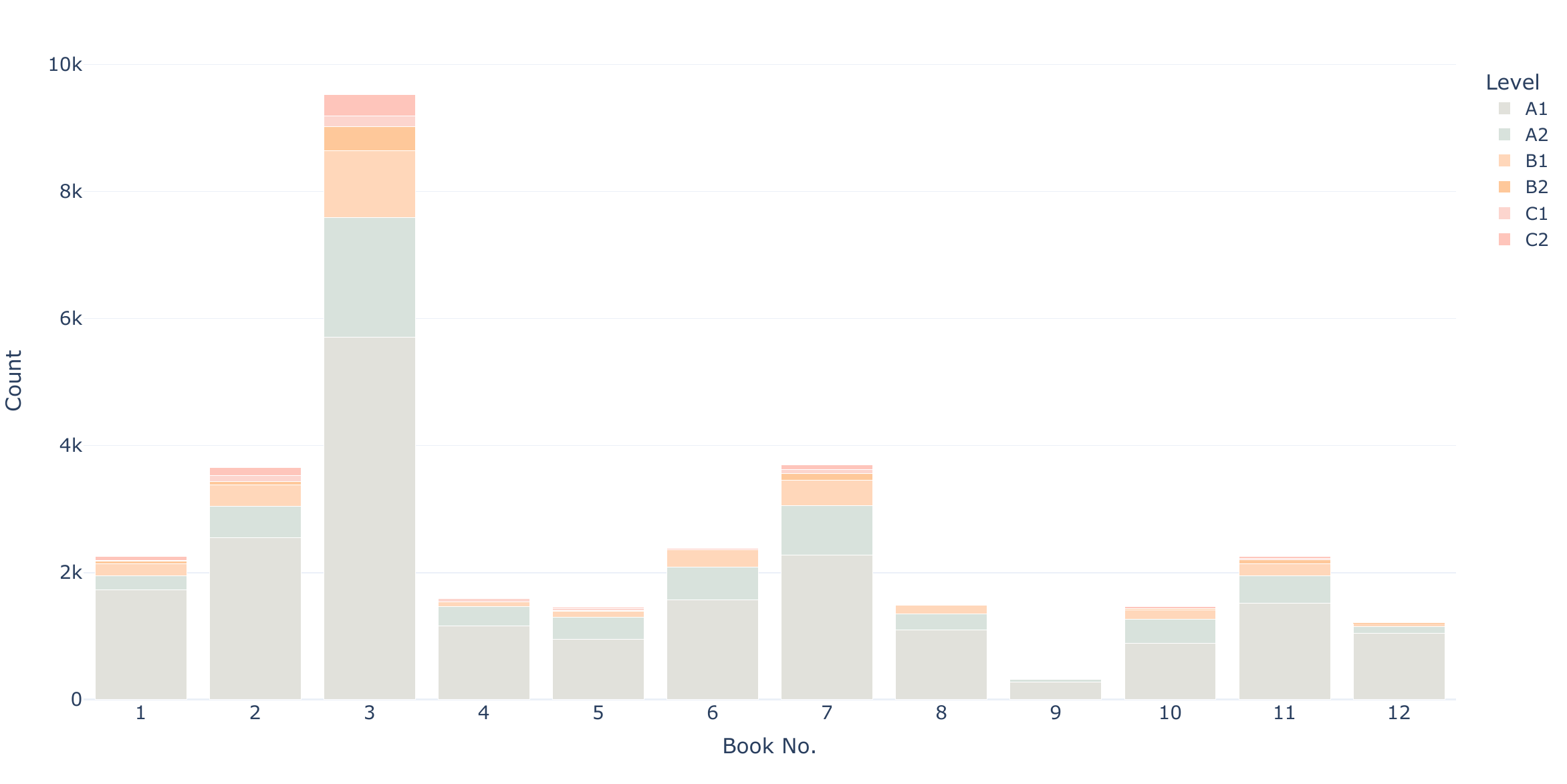}
  \caption{Distribution of the number of Python code constructs per textbook. Constructs are colored and stacked according to their competency level.}
  \label{fig:overview}
\end{figure*}
%%%%%%%%%%%%%%%%%%%%%%%%%%%%%%%%%%%%%%%%%%%%%%%%

%%%%%%%%%%%%%%%%%%%%%%%%%%%%%%%%%%%%%%%%%%%%%%%%
\begin{figure*}
  \centering
  \includegraphics[width=.9\textwidth]{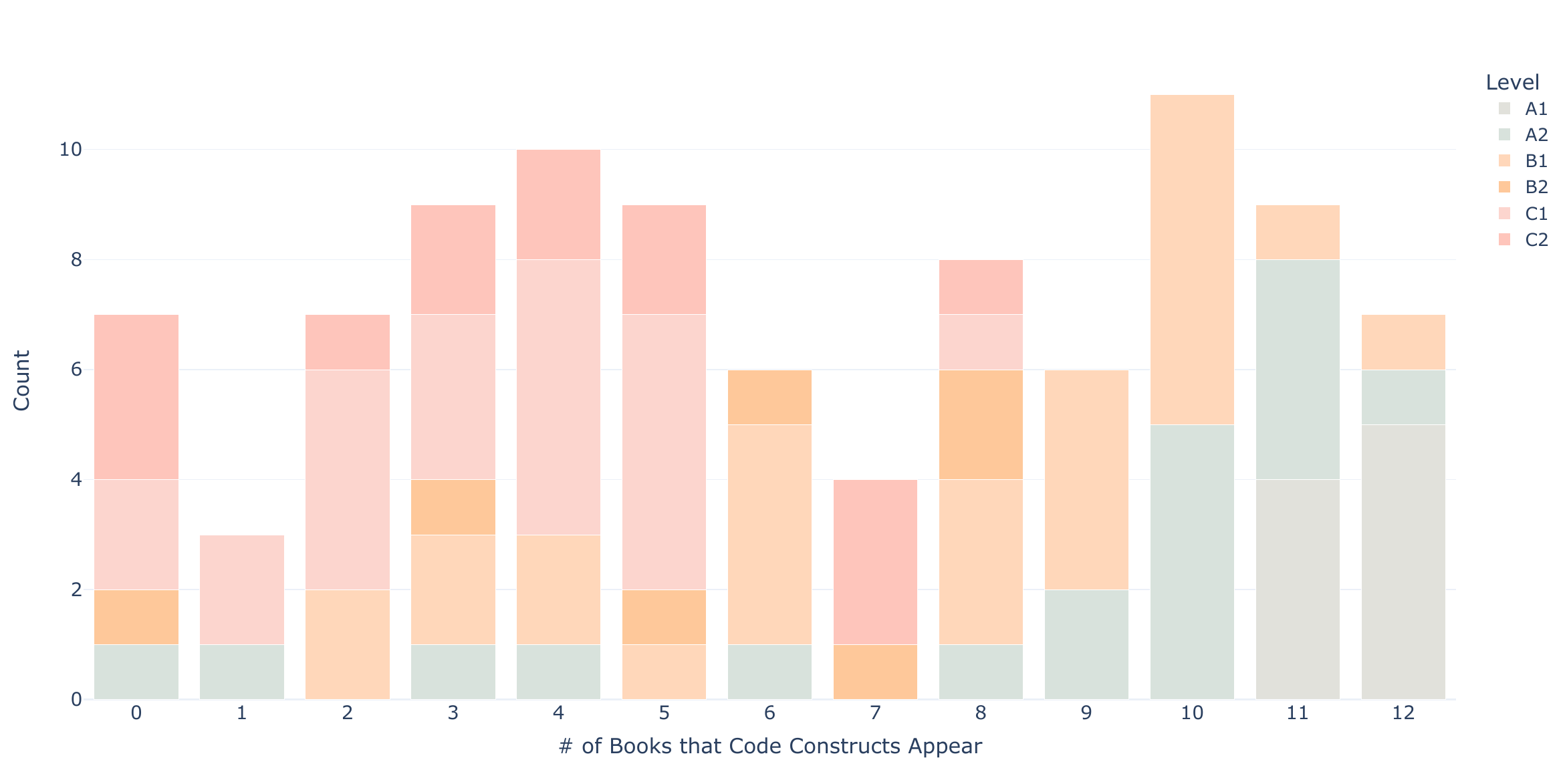}
  \caption{Distribution of the number of books in which Python constructs appear. Constructs are colored and stacked according to their competency level.}
  \label{fig:overview2}
\end{figure*}
%%%%%%%%%%%%%%%%%%%%%%%%%%%%%%%%%%%%%%%%%%%%%%%%

In this research question, we investigate if Python constructs appear in the textbooks, and if so, how often they do. We therefore first grab all the code constructs that have been extracted to see the proportion of each competency level in each book. Then, we analyze the frequency (in number of books in which they appear at least once) for each Python construct.

Figure~\ref{fig:overview} shows the total number of constructs, grouped by competency level, identified in each book. Each construct may appear more than once. A great variability in terms of the total number of constructs in the books can be observed: book 9 contains a few hundred, while book 3 is almost 10k. The other relevant issue is that the code that we have found in the books contains mostly elements of level A1, and then of A2, B1, and B2. Elements of C1 and C2 appear less frequently, and sometimes -contrary to what we expected- there are more elements of C2 than of C1. In any case, it is significant that there are books that have almost no C level elements. Even book number 9, aimed at children, has exclusively A1 elements.

Figure~\ref{fig:overview2} gives us a different perspective of the content in the textbooks. In it, we can visualize how many of the 12 books contain each of the 94 constructs. The colors indicate the competency level to which a particular construct belongs to. As can be seen in the figure, the elements of A1 appear in almost all the books: 6 constructs do so in all 12, while 5 in all books but one. As expected, in general terms, the frequency of the constructs decreases according to their difficulty level, with the bars turning pink (the tone of C1 and C2) as we get closer to the origin. Even so, there are several elements of level A2 and B1 that appear infrequently. For instance, \textit{nested tuples} (i.e., tuples of tuples, A2) do not appear in any book, and just one book contains a for loop with a \textit{tuple as iterator variable} (e.g., \textit{for (a, b) in iterable}, A2). Relative imports (B1) appear only in 2 books, and the combinations of using a \textit{list as an element of a dictionary} or using a \textit{dictionary as an element of a list} (both B1) only appear in 3 out of the 12 books. The other A2 constructs that appear infrequently are: in 3 books we find \textit{for loops iterating over a tuple} (e.g., \textit{for i in (1, 2, 3)}), in 4 books the \textit{writelines} function and in 6 books \textit{for loops iterating over lists} (e.g., \textit{for i in [1, 2, 3]}). If these constructs appear infrequently, it may be a good idea to i) recommend to include them in the textbooks or ii) reconsider them as a higher competency level (e.g., B2). We find that \textit{loops iterating over tuples} or \textit{loops iterating over lists} should be included in introductory textbooks as they help novices to better understand concepts on iteration (and iterables, etc.). They would strengthen the \textit{simple} for loop (e.g., \textit{for day in week}, A1) that is always present in textbooks. On the other hand, \textit{nested tuples}, \textit{writelines} and \textit{tuple as iterator variable} should be, in our opinion, reclassified to B1.

Table~\ref{tab:always-missing} provides information on the elements that appear in all the textbooks: 6 are from level A1, while there is one from A2 and B1, respectively. This makes us think that the level for these two elements (\textit{import} a function and \textit{simple while loop}) should be reconsidered to A1. On the other hand, we have those elements that do not appear in any book: 1 from level A2 (already discussed in the previous paragraph), one from B2 and three from C2. While this is expected for elements of C2 in an introductory book (in all three cases these are complex \textit{dictionary comprehensions}), or \textit{dictionary of lists} (B2). This seems to indicate that complex data structures have no place in introductory Python programming books, which makes sense.

\subsection{\rqthree}

%%%%%%%%%%%%%%%%%%%%%%%%%%%%%%%%%%%%%%%%%%%%%%%%
\begin{figure*}
  \centering
  \includegraphics[width=.9\textwidth]{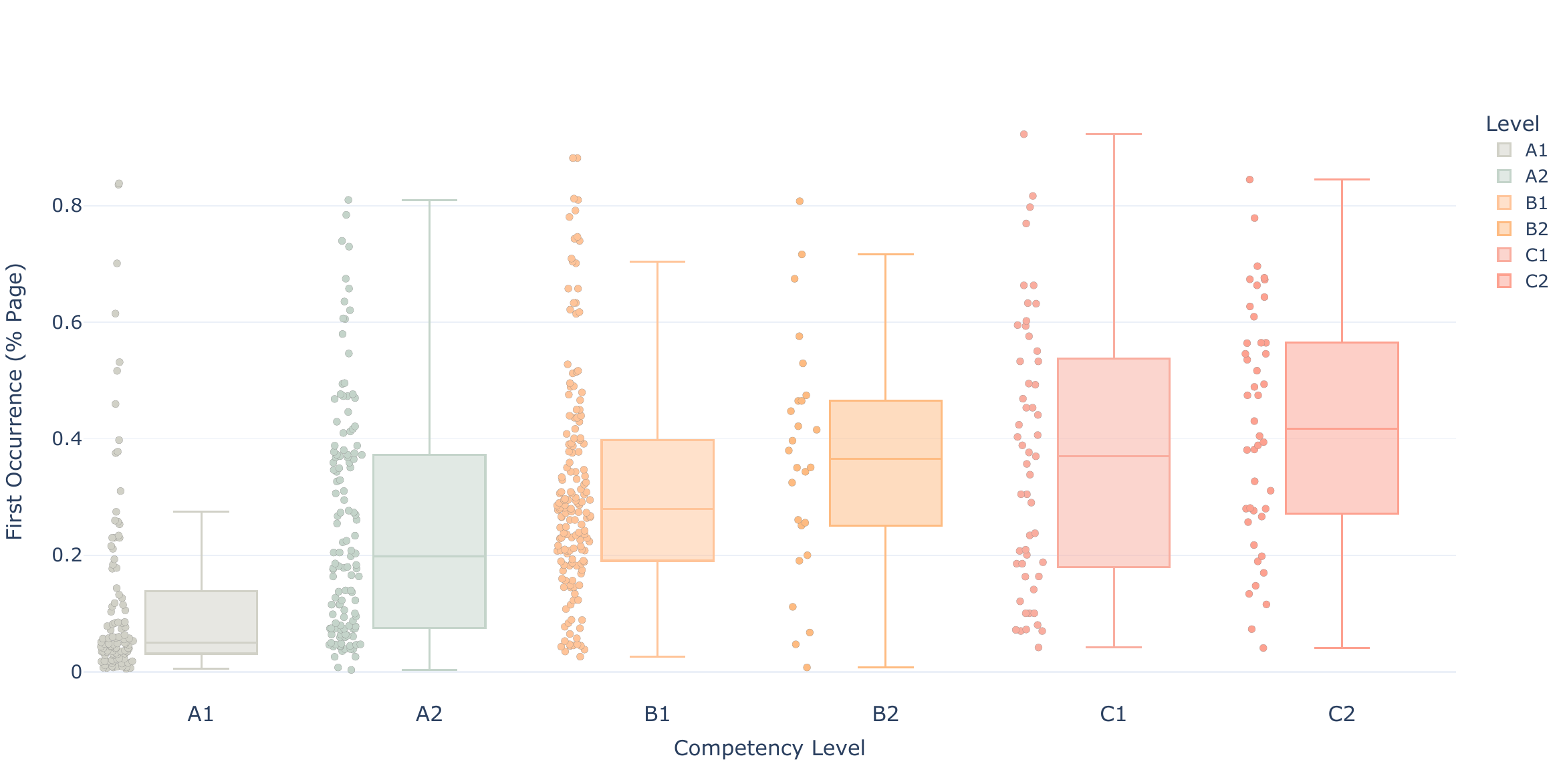}
  \caption{Distribution of code construct introduction in each level}
  \label{fig:rq3-sequenceofconstructs}
\end{figure*}
%%%%%%%%%%%%%%%%%%%%%%%%%%%%%%%%%%%%%%%%%%%%%%%%

In addition to whether Python constructs appear or not, an important question is where they are introduced (e.g., presented for the first time). As already noted, our assumption is that from a pedagogic point of view, textbooks should offer a path that begins by showing simple elements and progressively more complex ones. In other words, elements classified as A1 should appear at the beginning of a textbook, then those classified as A2, and so on.

Figure~\ref{fig:rq3-sequenceofconstructs} shows a first approximation to answer this RQ. We have taken the first appearance of each construct. Then, based on the page where it appears and according to the total number of pages in the book, we calculate the ratio where it is introduced. The results are very interesting, because although it is true that the six competency levels show a monotonically increasing median, some other aspects can be observed. Thus, there are elements of A1 that appear very late in the books. We have also found evidence of the contrary; there are elements of B2, C1, and C2 very early on. While the large point clouds for A1, A2, and B1 seem to indicate that in general terms (and not taking into account the outliers) a certain progression is followed, the point clouds of B2, C1, and C2 are more dispersed and seem almost identical, implying that for these levels the classification might have room for improvement. In other words, it seems that \textit{pycefr}'s proposal for assigning levels is, with some exceptions, relatively consistent with what we can find in textbooks for lower levels. However, for the higher levels, what we observe is that this allocation is not in line with what we have found in the textbooks. It should be noted that this visualization has some limitations. The most relevant one is that by showing the percentage of pages, we are normalizing all books. This might distort the results, as not all books are of the same siz

\begin{table}
    \centering
    \caption{List of most common and missing code constructs}
    \label{tab:always-missing}

    \begin{tabular}{p{2cm}cp{3cm}} % Adjust the width as needed
    \toprule
    &\textbf{Level}& \textbf{Code Construct} \\
    \toprule
    \textbf{Common} & A1& printfunc \\ &&simpleassign\\  &&assignwithsum\\&&simplelist\\ &&forsimple\\&& returnstatement \\&A2&importfunc\\  &B1&whilesimple \\
    \midrule
    \textbf{Missing} &A2& nestedtuple\\ &B2&nesteddictwithlist\\ &C2&dictcompwithifelse\\ &&dictcompwithif\\ &&nesteddictcomp\\
    \bottomrule
    \end{tabular}
%    \vspace{-10pt}
\end{table}

\begin{table}[]
\centering
\caption{Distance Scores for each textbook}
\label{tab:book-distance}
\begin{tabular}{ccc}
 \toprule
\textbf{Book No.} & \textbf{Distance} & \textbf{Relative Distance} \\ \toprule
1   & 54  &  1.10 \\
2   & 66 & 0.86 \\
3   & 92 & 1.08 \\
4   & 64  & 0.82 \\
6   & 32  & 0.68 \\
7   & 66  & 0.96 \\
8   & 26   & 0.76 \\
9   & 8    & 0.57 \\
10  & 38   & 0.88 \\
11  & 66 & 1.12   \\
12  & 34  & 0.79  \\\bottomrule 
\end{tabular}
\vspace{-8pt}
\end{table}

We have, therefore, delved deeper into this aspect using another approach: we have taken \textit{pycefr} as our \textit{ground truth} and present a measure of distance to quantitatively evaluate how \textit{far} the sequences in which Python constructs are presented in the introductory books are to this \emph{ground truth}. The procedure is as follows: for each textbook, we create a list with the Python constructs in the order they are introduced. The elements of the list are then replaced by their competency level as of \textit{pycefr}. This list is then compared with a list of the same size but \textit{perfectly} ordered, i.e., it will contain the same amount of elements of each level, but these elements will be ordered. In other words, if we had a sequence of six constructs in a book such as [A1, A1, A2, A1, B2, A2, C1], its \textit{perfect} sequence would be [A1, A1, A1, A2, A2, B2, C1]. 

We will then compute a distance metric between each pair of lists. We have chosen the weighted Levenshtein distance (WLD)~\cite{lhoussain2015adaptating}, as the \textit{vanilla} Levenshtein distance assumes a cost of 1 for all edit operations (i.e., all changes are of the same weight). The WLD, however, considers that changes may have different weights; it is used, for instance, for OCR correction, where substituting ‘0’ for ‘O’ should have a smaller cost than substituting ‘X’ for ‘O’. In our case, substituting A1 for A2 will have a smaller cost (1) than doing it for C2 (5).

Table~\ref{tab:book-distance} offers the results for the 12 books under consideration. The total distance is the sum of (weighted) edits for a textbook, and ranges from 8 to 92. The \textit{relative} distance is probably more valuable, given the different number of constructs the book contains. We calculate it by dividing the (total) distance by the number of constructs; this gives a relative measure of how far, in the mean, every construct is to where it belongs if we consider \textit{pycefr} as the \textit{ground truth}. Values now range from 0.57 for book 9 to 1.42 for book 4. In our opinion, this result offers evidence that the books do not follow \textit{pycefr}'s assignment; for most of the books, the relative distance is close to one level away in the mean. In the next RQ, we go deeper into this aspect.

\subsection{\rqfour}

\begin{table*}[]
\centering
\caption{Top 15 Python code constructs with the highest relative distance compared to their position in the \textit{perfect} sequence.}
\label{tab:code-distance}
\begin{tabular}{c|c|c|c|c}
 \toprule
\textbf{Pycefr level} & \textbf{Code Construct} & \textbf{Distance}&\textbf{Total Distance}   &\textbf{Relative Distance}                                                               \\ \toprule
        C2 & enumfunc & [4, 4, 3, 4, 3, 2, 3, 2] & 25 & 3.13 \\
        C2 & zip & [4, 4, 3, 3, 3, 0, 3] & 20 & 2.86 \\
        C2 & map & [4, 3, 0, 3, 0, 4, 4] & 18 & 2.57 \\
        C2 & listcompnested & [3, 3, 2, 1] & 9 & 2.25 \\
        C1 & simplelistcomp & [2, 2, 0, -1, 3, 3, 3, 3] & 17 & 2.13 \\
        C1 & importdbm & [2, 2] & 4 & 2 \\
        C1 & importre & [2, 0, 2, 3, 3] & 10 & 2 \\
        C1 & simpledictcomp & [2, 2] & 4 & 2 \\
        B1 & fromrelative & [-2, -2] & 4 & 2 \\
        A2 & fornested & [-1, -1, -1, -4, -1, -1, 1, -3] & 13 & 1.63 \\
        C2 & superfunc & [1, 3, 1, 3, 0] & 8 & 1.60 \\
        C1 & pickle & [2, 2, 0, 2, 2] & 8 & 1.60 \\
        B2 & \_\_class\_\_ & [1, 3, 1, 2, -1, 1, 2] & 11 & 1.57 \\
        C1 & struct & [0, 3] & 3 & 1.50 \\
        B1 & whilesimple & [2, 2, 2, -2, 2, 1, 1, 1, 1, 1, 1, 2] & 18 & 1.50 \\
        \bottomrule
\end{tabular}
\end{table*}

\begin{table}[]
\centering
\caption{Proportion of the disagreement}
\label{tab:stat-disagree}
\begin{tabular}{ccc}
 \toprule
\textbf{Dif Amount} & \textbf{Count} & \textbf{Percentage}                                                                 \\ \toprule
\rowcolor{pastelpink}
 -5   & 2  &  0.33                      \\
 \rowcolor{palepink}
-4   & 6 & 0.99 \\
 \rowcolor{mistyrose}
-3   & 18       & 2.98    \\
 \rowcolor{seashell}
-2   & 40   & 6.62     \\
 \rowcolor{snow}
-1   & 121     & 20.03       \\
0   & 243        & 40.23        \\
\rowcolor{snow}
1   & 101  & 16.72     \\
 \rowcolor{seashell}
2   & 39     & 6.46        \\
 \rowcolor{mistyrose}
3   & 26         & 4.30    \\
\rowcolor{palepink}
4  & 8   & 1.32    \\
\rowcolor{pastelpink}
5  & 0   & 0.00  \\
\midrule
\textbf{Total} & 604 & 100 \\
\bottomrule  

\end{tabular}
  \vspace{-8pt}
\end{table}

Given that we have found a certain disagreement between the assignments in \textit{pycefr} and the sequences in which Python constructs are presented in the textbooks, we aim to identify those Python constructs where the divergence is largest. Therefore, for every Python construct, we have compared its level (as per \textit{pycefr}) with the level of the element with the same index of the \textit{perfect} sequence for a given book. Thus, the distance for the Python constructs in the example we used in RQ3 would be [0, 0, +1, -1, +2, -2, 0]. A '0' means that the Python construct is where it should be; a '-2' means we have found the construct two levels above where \textit{pycefr} assigns it, while a '+2' is because the construct is located before than expected by \textit{pycefr}. This procedure is an approximation and is not free of noise, in particular for adjacent competency levels. However, to determine constructs that are very far from where they should be, as is our case, it is a good starting point.

Results can be found in Table~\ref{tab:stat-disagree}. We are analyzing the first appearance of 604 Python constructs in 12 books. Of these, 243 have been found at the level where they were expected to appear, which sums for slightly over 40\% of the sample. In total, almost 77\% of the constructs appear in their level, or one above or below. However, that also means that 23\% of the constructs appear further away. Of those, two appear 5 levels below where expected; this happens for C2 elements in A1 positions. We have not found the opposite, A1 constructs in C2 positions, but we have 4 cases of A1 where C1 was expected.

Having recognized the outliers, we will further analyze them on an individual basis. 
Table~\ref{tab:code-distance} lists the 15 Python constructs for which we have found the greatest divergence. In the first column, we can find their level, in the second the construct itself, and in the third a list with the differences between its level and the expected level for each book where the construct appears. Thus, for example, \textit{enumfunc} (i.e., the \textit{enumerate} function) appears in 8 books and is considered to be C2; according to the order in which it appears in the textbooks it should be A2 (difference of 4) three times, B1 (difference of 3) three times, and B2 (difference of 2) twice. The total distance (fourth column) is the sum of the absolute values of the differences. Since the total distance depends on the number of times the construct appears in the 12 books, a relative value is given in the last column (the sum of the absolute values divided by the number of books where it appears). Thus, the total distance for \textit{enumfunc} is 25, and its relative distance is 3.125.

We can observe that the most divergent elements in Table~\ref{tab:code-distance} correspond to elements considered as C that are introduced in the textbooks much earlier than expected. This is no surprise, since by the very definition of the metric these elements are the ones that can potentially reach the greatest distance, i.e., a B1 element can at most reach 2 or -3, while for a C2 element, the range goes from 0 to 5. If we analyze the elements of the list, we can observe the following issues:

\begin{itemize}[left=0pt, topsep=0pt]
   \item \textit{enumfunc} seems to have been erroneously assigned by \textit{pycefr} to C2 when in all the books in which it appears it is shown in positions A2-B2, with a predominance towards the lower levels (i.e., closer to A2 than to B2).
   \item \textit{map} and \textit{zip} are two functional programming constructs that allow you to perform advanced actions with lists. In many textbooks, they appear in positions that correspond to A2 and B1. Although we understand that C2 is perhaps too high, we also believe that its complexity does not correspond to a basic level (A). From our experience, we would put it as B2, but further evaluations should be performed on the fine-tuning.
   \item \textit{simpleListComp} and \textit{listCompNested} refer to list comprehensions and nested list comprehensions. In \textit{pycefr} they are classified as level C1 and C2, respectively. In the textbooks they appear, with some exceptions, much earlier. We understand that this makes sense, and that perhaps they could be reclassified as B1 for list comprehensions and, due to their added complexity, B2 for nested list comprehensions. A similar case is \textit{simpleDictComp}, which instead of C1 could be B1 or B2.
   \item \textit{importdbm} and \textit{importre} point to the use of functionality from the database and regular expression modules, respectively. They are now C1, but the input from the textbooks points out that B1 would be better suited. The constructs \textit{superfunc}, an object that represents the parent class, \textit{pickle}, functionality to convert a Python object hierarchy to a byte stream, \textit{\_\_class\_\_}, an attribute to show what class the object belongs to, are similar cases and could be classified at a lower level as well.
   \item \textit{fromrelative} (B1), functionality to import functions from other modules relative to the current location, and \textit{fornested} (A2), two nested for loops, are the first items in the list where \textit{pycefr} assigns lower levels than what we have identified in the books. Although we agree that \textit{fromrelative} could be B2, we do not find it suitable to be considered for proficiency (C1). With \textit{fornested}, either B1 or B2 would make more sense.
\end{itemize}

\begin{figure}
    \centering
    \begin{tcolorbox}[ width=8cm, height=7.2cm, colback=white, colframe=blue!50!black, boxrule=0.5pt, colbacktitle=gray, title={Python 3 for Absolute Beginners~\cite{hall2010python}, P.83-84}, fonttitle=\bfseries, top=4pt, bottom=2pt]
    The positional index and value can be both retrieved at once using \verb|enumerate()|, another built-in function.
    \begin{lstlisting}
    for i, value in enumerate(fruits):
        print i, value\end{lstlisting}
    \vspace{-0.55cm}
    To loop through two or more sequences at a time, entries can be paired with \verb|zip()|.
    \begin{lstlisting}
    for fru, veg in zip(fruits, vegetables):
        if fru < veg:
            print(fru, "are better then", veg)
        else:
            print(veg, "are better than", fru)\end{lstlisting}
    \vspace{1.0cm}
    \end{tcolorbox}
    \caption{zip and enumerate functions in book 4.}
    \label{fig:rq6-ex1}
      % \vspace{-8pt}
\end{figure}

\begin{figure}
    \centering
    \begin{tcolorbox}[ width=8cm, height=6.8cm, colback=white, colframe=blue!50!black, boxrule=0.5pt, colbacktitle=gray, title={A Python Book: Beginning Python, Advanced Python, and Python Exercises \cite{kuhlman2011python}, P.20}, fonttitle=\bfseries, top=2pt, bottom=2pt]
    Constructor for dictionaries \verb|dict()| can be used to create instances of the class dict. Some examples:
    \begin{lstlisting}
    dict({'one': 2, 'two': 3})
    dict({'one': 2, 'two': 3}.items())
    dict({'one': 2, 'two': 3}.iteritems())
    dict(zip(('one', 'two'), (2, 3)))
    dict([['two', 3], ['one', 2]])
    dict(one=2, two=3)
    dict([(['one', 'two'][i-2], i) for i in (2, 3)])\end{lstlisting}
    \vspace{-0.55cm}
    For operations on dictionaries, see http://docs.python.org/lib/typesmapping.html 
    \end{tcolorbox}
    \caption{zip function in book 11.}
    \label{fig:rq6-ex2}
\end{figure}

\begin{figure}
    \centering
    \begin{tcolorbox}[ width=8cm, height=13.5cm, colback=white, colframe=blue!50!black, boxrule=0.5pt, colbacktitle=gray, title={Learning Python, 5th Edition \cite{lutz2013learning}, P.111-113}, fonttitle=\bfseries, top=4pt, bottom=2pt]
    In addition to sequence operations and list methods, Python includes a more advanced operation known as a \verb|list comprehension expression|, which turns out to be a powerful way to process structures like our matrix. Suppose, for instance, that we need to extract the second column of our sample matrix. It’s easy to grab rows by simple indexing because the matrix is stored by rows, but it’s almost as easy to get a column with a \verb|list comprehension|:
    \begin{lstlisting}
    >>> col2 = [row[1] for row in M] >>> col2 # Collect the items in column 2
    [2, 5, 8]
    >>> M
    [[1, 2, 3], [4, 5, 6], [7, 8, 9]] # The matrix is unchanged\end{lstlisting}
    \vspace{-0.55cm}
    The \verb|map| built-in can do similar work, by generating the results of running items through a function, one at a time and on request. Like range, wrapping it in a list forces it to return all its values in Python 3.X; this isn’t needed in 2.X where map makes a list of results all at once instead, and is not needed in other contexts that iterate automatically, unless multiple scans or list-like behavior is also required:
    \begin{lstlisting}
    >>> list(map(sum, M)) # Map sum over items in M 
    [6, 15, 24]\end{lstlisting}
    \end{tcolorbox}
    \caption{\textit{map} function with list comprehensions in book 3.}
    \label{fig:rq6-ex3}
      \vspace{-8pt}
\end{figure}

\subsection{\rqfive}

In this RQ we analyze more in detail those Python constructs that are more divergent. For that reason, we have analyzed the code snippets as they appear in the textbooks, and have manually evaluated if their position makes sense, i.e., considering if learners are able to understand the Python construct at that time. This process has been done on an individual basis, snipped by snippet. We present and discuss those snippets where the divergence is larger in the paper, and point to the reproduction package for more cases.

Figure~\ref{fig:rq6-ex1} showcases examples of the \textit{enumerate} and \textit{zip} functions, both C2, in book 4. Both functions are introduced rather abruptly on pages 83 and 84 (out of a total of 300 pages in the book). Although we have argued before that probably these constructs should be categorized with lower levels, we find here explanations to be too concise, and doubt they could be followed by a beginner programmer right away. To exacerbate the situation, this is the unique reference to the \textit{zip} function within the entire book.

Figure~\ref{fig:rq6-ex2} demonstrates the first usage of the \textit{zip} function in book 11, in a code snippet alongside the introduction of the dictionary concept. Surprisingly, there is no accompanying explanation of the \textit{zip} function in the text. This example can be found very early (page 20 in a book of 278 pages), so we argue that a novice reader would not be able to understand it properly.

On the contrary, Figure~\ref{fig:rq6-ex3} serves as a commendable example of an effective explanation, facilitating a deeper understanding of the \textit{zip} the function. Despite this, a lingering question arises regarding the decision to introduce a \textit{list comprehension} and the \textit{map} function in the early stages of the book (pages 111-113 of a book with 1519 pages).

These examples underscore the importance of providing comprehensive explanations for the code presented, and not considering only the sequence in which concepts (and constructs) are introduced in a textbook.

\section{Lessons learned}
\label{sec:discussion}
After mining introductory Python programming textbooks, identifying the Python constructs, comparing the sequence in which they are introduced in the books with the levels proposed by \textit{pycefr}, and analyzing in detail constructs where the divergence is largest, we can draw some lessons learned. In particular:

\begin{itemize}[]
    \item \textbf{\textit{Python constructs in lower levels (i.e., A2) that are seldom introduced in the textbooks.}} We recommend some of these constructs to be included in the textbooks (e.g., iterating over lists or tuples) while others should be reclassified to other, more advanced levels (e.g., nested tuples or \textit{writelines}).
    \item The \textit{\textbf{simple while loop (A2) and import a function (B1) seem to be more appropriately assigned to A1}}, as they appear in the early stages of all 12 books analyzed. 
    \item Although in 80\% of the cases, the assignment of \textit{pycefr} seems to be close to the sequences found in books, there is still a relatively high percentage of divergence.
    \item \textbf{\textit{The divergence seems to be more frequent for complex constructs (B2, C1, and C2) than for basic ones (A1, A2, and B1).}} This makes us think that the assignment is much clearer in basic elements than in complex elements. It seems, therefore, that the books follow a more or less homogeneous sequence in the basic levels, and that they diverge to a greater extent when the more advanced elements are introduced.
    \item \textbf{\textit{Extreme divergence (i.e., constructs that are C2 for \textit{pycefr} but are presented early in the textbooks) is not uncommon.}} We have discussed some of these cases individually and tend to recommend to assign lower levels to those constructs.
    \item We identified corner cases where \textit{\textbf{advanced constructs are sometimes out of place.}} Thus, some functions have caught our attention, especially \textit{map} and \textit{zip}, which appear in our opinion too early in the books. In RQ5 we have analyzed these particular cases by looking at the snippets in the books, and have found that sometimes they are not presented exhaustively. We affirm, therefore, that from a didactic point of view, it would be advisable to introduce these constructs earlier as in \textit{pycefr} (C2), but later than in the textbooks.
    \item There are a \textbf{\textit{few constructs that \textit{pycefr} classifies with lower levels, but consistently appear later in the books.}} In particular, among the 15 most extreme cases, we have identified \textit{fornested} and \textit{fromrelative}.
\end{itemize}

\section{Threats to validity}
\label{sec:threats}

According to Wohlin et al.~\cite{wohlin2012experimentation}, there are four types of threats to validity in software engineering:

Construct validity refers to the degree to which a study measures what it intends to measure. One aspect that might affect our study is poor measurement tools; the regular expressions that we have used might be buggy. To deal with this threat, we have evaluated the performance of our regular expressions in the RQ1. However, since the number of expressions is 94 and the sample size of our evaluation was 380 out of 31,333 cases, our evaluation process might have room for improvement.

Internal validity refers to the degree to which a study can establish a causal relationship between the independent and dependent variables. We could have selection bias, which might have occurred because the textbooks we have found are not representative of the population of interest. The number of analyzed textbooks is 12, but there may be many more of them. In addition, for the analysis we do not read the books, but just performed mining with regular expressions; thus, there might be information that we have missed. We do not have a \textit{ground truth} for the assignment of competency levels to Python constructs. We have used the proposal by the \textit{pycefr} authors and compared it to the sequences found in 12 books, but still we would require more information to create a solid assignment. We have used the concept of \textit{perfect} sequence as the basis to measure distances and identify divergence in level assignment; the \textit{perfect} sequence can be noisy and may not discriminate between levels that are close. That is why we have focused on those constructs that are more divergent.

External validity refers to the degree to which the results of a study can be generalized to other populations, settings, or times. We cannot claim that our results are valid for any other programming language. Since we were focusing on the Python programming language, textbooks are especially thought for learning Python. Even if the programming concepts in other programming languages might be similar, it is to be seen if our findings are applicable there too.. To comply with submission requirements, we have anonymized the authors' identities

Conclusion validity refers to the degree to which the conclusions of a study are justified by the data. In this regard, we might have failed to consider alternative explanations for the results.

\section{Conclusion and Future Outlook}
\label{sec:conclusions}

In this paper, we have evaluated \textit{pycefr}'s assignments of almost 100 Python constructs to competency levels.
To do this, we have systematically analyzed introductory textbooks by comparing the sequences found in the textbooks with the proficiency levels proposed by the \textit{pycefr} tool, which are based on a single textbook and the opinion of developers and experts.
We have substantiated that while assignments are generally aligned in the majority of cases, there exists a discernible amount of disagreement. We have identified several cases where this disagreement is large, studying them in detail. 
We have also been able to find some cases of constructs that are taught too soon and in a very superficial way in the books, giving not only rise to learners not understanding the concept (and the code) and having doubts, but also demonstrating that some caution must be exercised when considering textbooks as the \emph{ground truth} for sequence identification. 
In other words, the order of presentation of Python constructs in books can also be noisy; if we want to use the method proposed in this paper to obtain a \emph{complete} assignment of Python constructs to levels, these inconsistencies have to be identified and filtered.

In the realm of software engineering, the importance of having tools to accurately assess a developer's proficiency in programming languages—and the competency required to comprehend, rectify, and enhance software systems—cannot be overstated. This need becomes even more critical with the emergence of technologies like Large Language Models (LLMs) that have the capability to generate code based on natural language instructions. The ability for developers to understand code is essential not only for ensuring and auditing the quality of software but also for drawing inspiration from existing code to suggest improvements.
Our vision extends to leveraging developer competency levels as input parameters for LLMs, thereby ensuring that the code generated is within the developer's comprehension and capability to manage, correct, and ultimately, take responsibility for. This approach underscores the synergy between human expertise and artificial intelligence in software development, emphasizing the vital role of developer competency in steering the responsible and effective use of LLMs in coding practices.

This paper represents a preliminary step in this direction. 
Given the importance of code proficiency, although it does not provide a comprehensive mapping of Python constructs to proficiency levels, it demonstrates the potential for achieving this through textbook analysis. The finding underscores the need to develop more precise methods for determining code proficiency for each code construct.  Future steps could involve correlating these findings with actual developer activities over time leading to new research directions. 
% We aspire that this work inspires other researchers to explore this avenue in the near future.

\section*{Data Availability}
The dataset with the code construct extraction and the analysis for each RQ are available anonymized in Zenodo, at \url{https://zenodo.org/doi/10.5281/zenodo.10149964}.

\section*{Acknowledgement}
This work is supported by the JSPS KAKENHI Grant Number JP20H05706 and the Spanish Ministry of Science, Innovation, and Universities under the Excellence Network AI4Software (Red2022-134647-T) and through the Dependentium project (PID2022-139551NB-I00).

\bibliographystyle{IEEEtran}
\bibliography{references}

\end{document}